\documentclass[a4paper, 10pt, oneside, onecolumn, notitlepage, final]{article}
\usepackage[text={160mm, 240mm}, left=25mm, vmarginratio=1:1]{geometry}
\usepackage{graphicx}
\usepackage{amssymb, amsfonts, amsmath, amsthm, mathtools, mathrsfs}
\usepackage{upgreek}
\usepackage{array}
\usepackage{bigstrut}
\usepackage[usenames, dvipsnames]{xcolor}
\usepackage[colorlinks, linkcolor=blue, anchorcolor=red, citecolor=blue]{hyperref}
\usepackage[backend=biber, sorting=none, maxbibnames=3, minbibnames=3, style=numeric-comp]{biblatex}
\usepackage{braket} % for bra, ket and set
\usepackage{authblk} % required for affiliation
\usepackage{orcidlink}
\usepackage[ruled, vlined]{algorithm2e}
\usepackage{abstract}
\usepackage{titlesec}
\usepackage[T1]{fontenc}
\usepackage[utf8]{inputenc}
\usepackage{lmodern}
\usepackage[english]{babel}

\titleformat{\section}{\large\bfseries\centering}{\thesection}{0.5 em}{}
\titleformat{\subsection}{\normalsize\bfseries\centering}{\thesubsection}{0.5 em}{}

\begin{document}

\title{\bf Generic Method for Integrating Lindblad Master Equations}

\author[1]{\normalsize Jiayin Gu\orcidlink{0000-0002-9868-8186}\thanks{\texttt{gujiayin@njnu.edu.cn}}}
\author[2,3]{\normalsize Fan Zhang\orcidlink{0000-0002-7466-6898}\thanks{\texttt{van314159@pku.edu.cn}}}

\affil[1]{\normalsize School of Physics and Technology, Nanjing Normal University, Nanjing 210023, China}
\affil[2]{\normalsize School of Physics, Peking University, Beijing 100871, China}
\affil[3]{\normalsize RIKEN Center for Emergent Matter Science, RIKEN, Saitama, 351-0198, Japan}

\date{}
\maketitle

%\vspace{-1xm}

\begin{abstract}
The time evolution of Markovian open quantum systems is governed by Lindblad master equations, whose solution can be formally written as the Lindbladian exponential acting on the initial density matrix. By expanding this Lindbladian exponential into the Taylor series, we propose a generic method for integrating Lindblad master equations. In this method, the series is truncated, retaining a finite number of terms, and the iterative actions of Lindbladian on the density matrix follow the corresponding master equation. Our method offers significant improvements in numerical efficiencies both in memory cost and computation time, especially for systems with many degrees of freedom. Moreover, our proposed method can be integrated seamlessly with tensor networks. Two illustrative examples, a two-level system exhibiting damped Rabi oscillations and a driven dissipative Heisenberg chain, are used to demonstrate the validity of our method. The superiority of our method is benchmarked with detailed performance tests.
\end{abstract}

\section{Introduction}

Almost all realistic quantum systems are unavoidably open, constantly interacting with their environments in the form of decoherence and/or heat exchange~\cite{Breuer_2002, Rivas_2012}. These interactions significantly alter the system's dynamics, leading to the emergence of unique features that are absent in closed systems. For example, energy dissipation generally drives the system toward a steady state where the growth of bipartite entanglement is inhibited~\cite{Aolita_RepProgPhys_2015}. Decoherence, traditionally viewed as a detrimental effect on coherent state manipulations, has also been leveraged and engineered for the preparation of highly non-trivial quantum states that are crucial for various applications in quantum computing and quantum simulation~\cite{Diehl_NatPhys_2008, Verstraete_NatPhys_2009, Weimer_NatPhys_2010, Georgescu_RevModPhys_2014, Takahashi_ProcJpnAcadSerB_2022, Daley_NatRevPhys_2023}. When the system-environment coupling is weak, the dynamics of an open quantum system is governed by the Lindblad master equation~\cite{Lindblad_CommunMathPhys_1976, Gorini_JMathPhys_1976, Manzano_AIPAdv_2020}, $\dot{\rho}=\mathcal{L}(\rho)$, where $\rho$ is the density matrix and $\mathcal{L}$, called the "Lindbladian," is a non-Hermitian superoperator acting linearly on $\rho$. While there are few cases where the Lindblad master equation can be exactly solved, such as the boundary-driven XXZ model~\cite{Prosen_PhysRevLett_2011a} and Bose-Hubbard model~\cite{Nakagawa_PhysRevLett_2021}, in most cases we have to resort to numerical methods.

One widely adopted method is the so-called vectorization method, which recasts the Lindblad master equation into the familiar matrix-vector form by concatenating the rows or columns of the density matrix into a single vector, and constructing the Lindbladian as a big matrix. As a result, the new state space is expanded. Alternatively, the quantum jump method, also known as the Monte Carlo wave function (MCWF) method, unravels the ensemble-averaged evolution in terms of individual quantum trajectories~\cite{Carmichael_1993, Wiseman_2009, Daley_AdvPhys_2014, Plenio_RevModPhys_1998}. In this method, we only need to propagate pure states that can be represented as vectors. The expense to pay is having to sample over many realizations. Also necessarily mentioned are the solvers for ordinary differential equations (ODEs), such as the canonical fourth-order Runge-Kutta method (RK4). These solvers are widely used in numerical packages, e.g., QuantumOptics.jl~\cite{Kramer_ComputPhysCommun_2018}, to integrate Lindblad master equations. However, these solvers are step-sensitive in terms of the numerical accuracy. For example, the overall integration error of the RK4 is on the order ${\cal O}(\delta t^4)$, where $\delta t$ is the time step. In contrast, the vectorization method is numerically exact. It is certainly possible to use Runge-Kutta methods of higher orders. But the intrinsic problem with Runge-Kutta methods is that the coefficients entering such methods are not uniquely determined. This arbitrariness gets more severe when the Runge-Kutta methods are of higher orders. As such, it adds complexity when numerically implementing Runge-Kutta methods.

As with their closed counterparts, exact numerical calculations for open many-body quantum systems are always hindered by the curse of dimensionality. Tensor networks have emerged as powerful tools for studying strongly correlated quantum many-body systems~\cite{Verstraete_AdvPhys_2008, Orus_NatRevPhys_2019, Cirac_RevModPhys_2021, Banuls_AnnuRevCondensMatterPhys_2023, Xiang_2023}. These techniques have also been extended to open systems using the vectorization method~\cite{Nakano_PhysRevE_2021, Weimer_RevModPhys_2021}. However, the vectorization method enlarges the matrix representation of the Lindbladian, leading to substantial redundancy that contradicts the spirit of tensor networks. This results in significant computational overhead, especially for large Hilbert spaces. Here, we propose a generic method for integrating the Lindblad master equation that circumvents the dimension-inflation issue inherent in the vectorization method. In this method, the Lindbladian exponential is expanded into a truncated Taylor series. Each term involves iterative actions of the Lindbladian on the initial density matrix, following the corresponding master equation. Compared to the vectorization method, our method is superior in numerical efficiencies both in memory cost and computational time, particularly for systems with many degrees of freedom.

This paper is organized as follows. We first review the basics of Lindblad master equations and the commonly used vectorization method to integrate them in Sec.~\ref{LindbladEquations}. Then, we give a brief introduction to common methods for computing matrix exponential in Sec.~\ref{MatrixExponential}, where it is clear that the computation of matrix exponential boils down to matrix multiplication. In Sec.~\ref{ProposedMethod}, we propose a generic method for integrating Lindblad master equations and clarify its advantages over other integrators. The validity of the proposed method is demonstrated in Sec.~\ref{Examples} with two examples: a two-level system exhibiting damped Rabi oscillations and a driven dissipative Heisenberg chain. In this section, the proposed method is also shown to integrate seamlessly with tensor networks. In Sec.~\ref{PerformanceTests}, the superiority in performance is shown through detailed performance tests. We finalize with conclusion and discussion in Sec.~\ref{Conclusion}.

\section{Lindblad Master Equations}\label{LindbladEquations}

Under the Born-Markov approximation, the dynamics of an open quantum system is described by the Lindblad master equation, reading
\begin{align}
\frac{{\rm d}\rho}{{\rm d}t}={\cal L}\rho=\frac{\rm i}{\hbar}[\rho,H]+\sum_i\left(L_i\rho L_i^{\dagger}-\frac{1}{2}\{L_i^{\dagger}L_i,\rho\}\right) \text{,} \label{eq_Lindblad}
\end{align}
where $[A,B]=AB-BA$ and $\{A, B\}=AB+BA$ are respectively the commutator and anticommutator of two operators, $H$ is the system Hamiltonian describing the unitary aspect of the dynamics, and \{$L_i$\} are a set of Lindblad operators describing the nonunitary (dissipative) part of the dynamics. The index $i=1,2,\cdots$ identifies the type of quantum jump, e.g., the removal (injection) of particles from (into) the open system. The formal solution of the Lindblad master equation~(\ref{eq_Lindblad}) is given by
\begin{align}
\rho(t)={\rm e}^{{\cal L}t}\rho(0) \text{,}
\end{align}
whose structure is similar to that of Schr\"odinger equation or classical master equation.

\par To integrate the Lindblad master equation~(\ref{eq_Lindblad}), the vectorization method is commonly used. The mathematical trick rooted in this method is the so-called "Choi isomorphism." It states that the coefficients of a matrix can be rewritten as those of a vector. Specifically, it consists of stacking the columns of $\rho=\sum_{m,n}\rho_{mn}\ket{m}\!\bra{n}$ into a single column vector $\vert\rho\rangle\!\rangle$ (column-wise vectorization with the first column on the top and the last column on the bottom)~\footnote{This map is a linear isometry between the $d\times d$ Liouville space of $\rho$ and the $d^2$ Hilbert space $|\rho\rangle\!\rangle$. It preserves norms with the correspondence between Frobenius norm and spectral norm, $\sqrt{{\rm Tr}\left(\rho^{\dagger}\rho\right)}=\sqrt{\langle\!\langle\rho|\rho\rangle\!\rangle}$.}. Correspondingly, the Lindbladian ${\cal L}$ in the matrix form is constructed as~\footnote{For any matrices $A$, $B$, and $X$, we have $|AXB\rangle\!\rangle=(I\otimes A)|XB\rangle\!\rangle=(I\otimes A)(B^{\rm T}\otimes I)|X\rangle\!\rangle=(B^{\rm T}\otimes A)|X\rangle\!\rangle$ for the column-wise vectorization.}
\begin{align}
{\cal L}= & -\frac{\rm i}{\hbar}\left(I\otimes H-H^{\rm T}\otimes I\right) \nonumber \\
+ & \sum_i\left[L_i^*\otimes L_i-\frac{1}{2}\left(I\otimes L_i^{\dagger}L_i+(L_i^{\dagger}L_i)^{\rm T}\otimes I\right)\right] \text{,}
\end{align}
where $I$ represents the identity matrix matching the dimension of $H$, $L_i^*$ denotes the complex conjugate of $L_i$, and $\otimes$ stands for the tensor product. A similar procedure exists for row-wise vectorization. Because $\rho$ is a $d\times d$ matrix, $|\rho\rangle\!\rangle$ is a $d^2$-dimensional vector, and ${\cal L}$ becomes a $d^2\times d^2$ matrix. After computing ${\rm e}^{{\cal L}t}|\rho\rangle\!\rangle$, the resulting vector can be converted back into matrix form to obtain the density matrix $\rho(t)$ at time $t$.

\section{Matrix Exponential}\label{MatrixExponential}

The computation of the matrix exponential is crucial for the subsequent analysis of numerical complexities. For this purpose, let us now delve into the details of how the matrix exponential is computed. We refer to Ref.~\cite{Moler_SIAMRev_2003}, where nineteen dubious ways to compute the exponential of a matrix are compiled, including methods through, e.g., approximation theory, differential equations, and matrix eigenvalues. Here, we focus on two methods based on approximation theory. The first one to mention is that by definition. For each complex matrix $M$, the exponential function is defined by the following Taylor series expansion:
\begin{align}
{\rm e}^M=\sum_{k=0}^{\infty}\frac{M^k}{k!}=1+M+\frac{M^2}{2!}+\frac{M^3}{3!}+\cdots \text{,} \label{eq_taylor1}
\end{align}
which converges for all complex matrices of any finite dimension. The function \verb"gsl_linalg_exponential_ss" in the GNU Scientific Library (GSL) implements the matrix exponential in this way~\cite{GSL_manual_2021}. The second and also more widely adopted one is by representing the matrix exponential as Pad\'e approximants~\cite{Baker_1975},
\begin{align}
{\rm e}^M & \approx R_{m,n}(M)=\frac{P_m(M)}{Q_n(M)} \text{,} \label{eq_pade}
\end{align}
where $P_m(M)$ and $Q_n(M)$ are polynomials of degrees $m$ and $n$, respectively. The Pad\'e approximants can be thought of as generalizations of the Taylor series; when the denominator takes $Q_0(M)=1$, the Pad\'e approximants reduce to the Taylor series. For numerical calculations in practice, the diagonal Pad\'e approximants ($m=n$) are more favored than off-diagonal ones ($m\neq n$), and in this case, the polynomials are explicitly given by
\begin{align} 
& P_m(M)=\sum_{k=0}^m\frac{(2m-k)!m!}{(2m)!(m-k)!}\frac{M^k}{k!} \text{,} \\
& Q_m(M)=P_m(-M) \text{.}
\end{align}
The diagonal Pad\'e approximants are believed to give more accurate results than with truncated Taylor series, and are thus widely implemented, e.g., in \texttt{expm} function in \textsc{scipy}~\cite{SciPy_manual_2022}, or in \texttt{exp} function in \textsc{julia}~\cite{Bezanson_SIAMRev_2017}. The two methods are often combined with the scaling and squaring technique to improve the accuracy. The technique scales the matrix by a power of 2 to reduce the norm, computes a truncated Taylor series or Pad\'e approximants to the scaled matrix exponential, and then repeatedly squares to undo the effect of the scaling,
\begin{align}
{\rm e}^M=\left({\rm e}^{M/2^s}\right)^{2^s} \text{.}
\end{align}
This technique efficiently balances the need for accuracy with computational feasibility, making it a popular choice for computing matrix exponentials in scientific computing. A catalog of software for matrix functions, including matrix exponential, is provided in Ref.~\cite{Higham_unpublished_2020}. For more comprehensive reference on matrix exponential, readers are directed to Ref.~\cite{Higham_2008}.

\begin{table*}
\caption{Comparison between the integrators in terms of the spatial and temporal complexities.}
\begin{center}
\begin{tabular}{>{\centering\arraybackslash}m{4.0cm}|>{\centering\arraybackslash}m{4.0cm}|>{\centering\arraybackslash}m{4.0cm}}
\hline
\hline
& Complexity in space & Complexity in time \bigstrut \\ \hline
vectorization method \#1 & ${\cal O}(d^4)$ & ${\cal O}(d^6)$ \bigstrut \\ \hline
vectorization method \#2 & ${\cal O}(d^4)$ & ${\cal O}(d^4)$ \bigstrut \\ \hline
Taylor series method & ${\cal O}(d^2)$ & ${\cal O}(d^3)$ \bigstrut \\ \hline
\hline
\end{tabular}
\end{center}
\label{tab_comparison}
\end{table*}

\begin{figure}
\centering
\begin{minipage}[t]{0.6\hsize}
\resizebox{1.0\hsize}{!}{\includegraphics{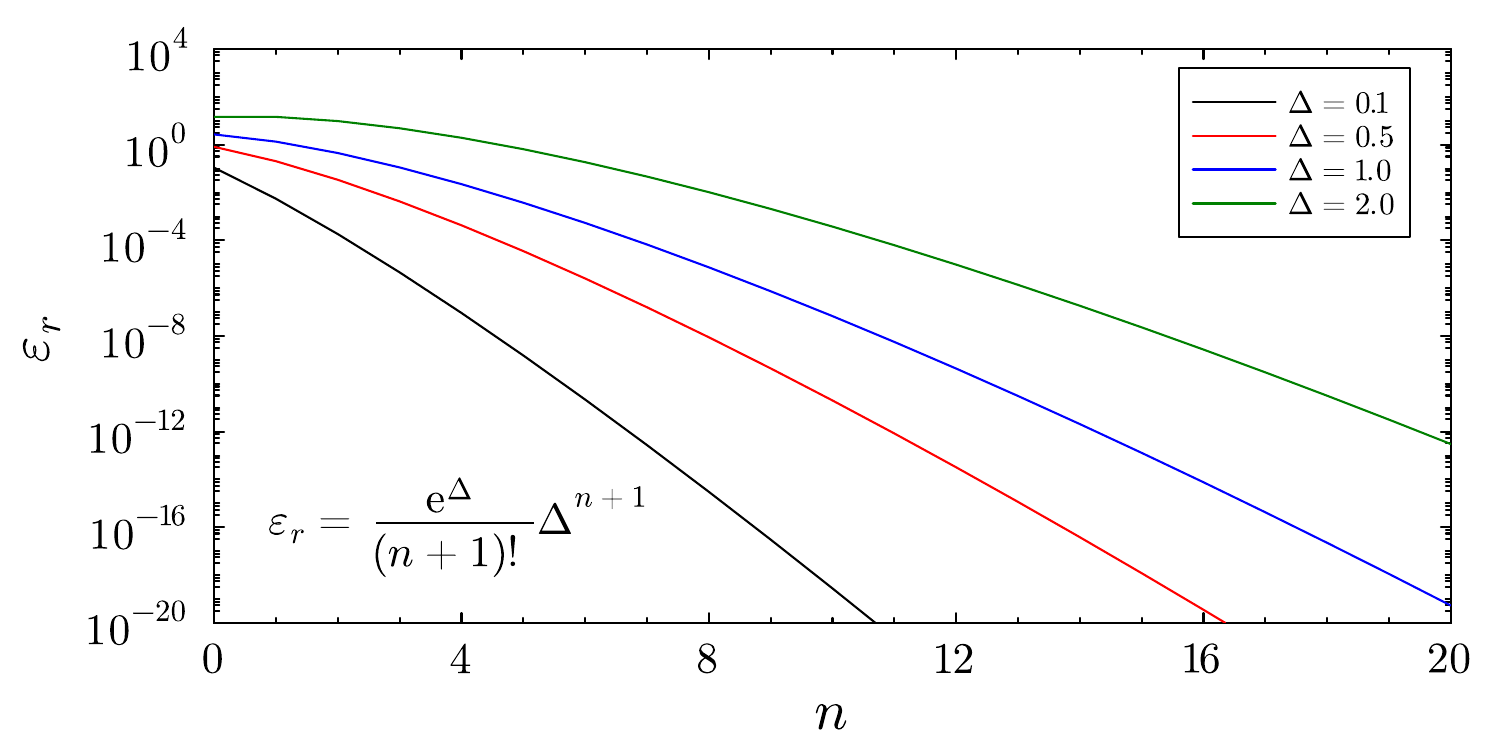}}
\end{minipage}
\caption{The decaying behavior of the relative error with the number of retained terms in the Taylor series method. The parameter $\Delta$ is given in terms of the Lindbladian norm and the time: $\Delta=||{\cal L}||t$.}
\label{fig_error}
\end{figure}

\section{Proposed Method}\label{ProposedMethod}

Now we analyze the numerical complexities involved in integrating the Lindblad master equation~(\ref{eq_Lindblad}). In the vectorization method, the numerical complexity in space depends on the matrix size, i.e., Lindbladian ${\cal L}$, scaling as ${\cal O}(d^4)$. We define the multiplication of two complex numbers as a unit of computation. If we compute the full Lindbladian exponential ${\rm e}^{{\cal L}t}$, which boils down to matrix-matrix product, then the numerical complexity in time is ${\cal O}(d^6)$. Subsequently, this method with full Lindbladian exponential is called vectorization method \#1. Alternatively, we can expand the Lindbladian exponential in Taylor series,
\begin{align}
\rho(t)={\rm e}^{{\cal L}t}\rho(0)=\sum_{k=0}^{\infty}\frac{t^k}{k!}{\cal L}^k\rho(0) \text{,} \label{eq_taylor2}
\end{align}
indicating that we can iteratively apply the matrix form of ${\cal L}$ on the initial $\rho(0)$ of the vectorized form. In this way, each term can be evaluated, and the sum of them gives the desired result. The numerical complexity in time scales as ${\cal O}(d^4)$. We call this method with Taylor series vectorization method \#2. These numerical complexities grow rapidly with the dimension of Hilbert space. To alleviate this issue, we propose a generic method to integrate the Lindblad master equation~(\ref{eq_Lindblad}). Starting from the Taylor series expansion~(\ref{eq_taylor2}), the Lindblad master equation can be integrated directly without vectorizing the density matrix. The iterative action of ${\cal L}$ on $\rho(0)$ follows the rule according to the Lindblad master equation. It can be easily checked that the trace is preserved as ${\rm Tr}[{\cal L}^k\rho]=0$ for $k=1,2,\cdots$. Considering that the action of ${\cal L}$ on $\rho$ only involves the matrix multiplication of the Hamiltonian $H$ and Lindblad operators $\{L_i\}$ on both sides, and all of these are $d\times d$ matrices, the numerical complexities in space and time scale as ${\cal O}(d^2)$ and ${\cal O}(d^3)$, respectively. These are largely reduced compared with those of the vectorization method, offering a huge advantage in numerical efficiency, especially when the dimension of Hilbert space is very large. Here, we have assumed that the number of Lindblad operators does not increase with the dimension of the system, such as the case of the Heisenberg chain considered in Sec.~\ref{Examples}. This assumption is quite natural for a lot of quantum many-body systems in which decoherences are induced locally at their interactions with the surrounding environment. The comparison in numerical complexities between the Taylor series method and the vectorization method is summarized in Table~\ref{tab_comparison}. Here, we point out that the Runge-Kutta methods have the same numerical complexities as the proposed method. See Appendix~\ref{RK4} for the application of RK4 to integrate the Lindblad master equation~(\ref{eq_Lindblad}). However, the numerical accuracy that heavily depends on the time step in Runge-Kutta methods is an issue. We will discuss this later in the text.

\begin{algorithm}
\caption{Compute $\rho(t+\delta t)={\rm e}^{{\cal L}\delta t}\rho(t)$ according to the Taylor series method.}\label{algorithm}
\KwIn{density matrix $\rho(t)$;}
\KwIn{Hamiltonian $H$;}
\KwIn{Lindblad operators $\{L_i\}\equiv\{L_1,\cdots,L_s\}$;}
\KwIn{time step $\delta t$;}
\KwIn{number of retained terms in Taylor series $n$;}
$\rho_m\gets\rho(t)$\;
$\rho_{\rm sum}\gets\rho(t)$\;
\For{$k$ in $1,\cdots,n$}
{
	$\rho_{\rm temp}\gets-{\rm i}/\hbar(H\rho_m-\rho_mH)$\;
	\For{$L_i$ in $L_1,\cdots,L_s$}
	{
		$\rho_{\rm temp}\gets\rho_{\rm temp}+L_i\rho_m L_i^{\dagger}$\;
		$\rho_{\rm temp}\gets\rho_{\rm temp}-\frac{1}{2}\left(L_i^{\dagger}L_i\rho_m+\rho_mL_iL_i^{\dagger}\right)$\;
	}
	$\rho_m\gets\rho_{\rm temp}$\;
	$\rho_{\rm sum}\gets\rho_{\rm sum}+\rho_m(\delta t)^k/k!$\;
}
\KwOut{$\rho(t+\delta t)\equiv\rho_{\rm sum}$;}
\end{algorithm}

\par In practical calculations, the Taylor series~(\ref{eq_taylor2}) is truncated up to a finite number of terms. The time $t$ in each step takes a small quantity so that the truncated Taylor series represents a sufficiently good approximation to the matrix exponential. The number of terms in the truncated Taylor series is a well-controled parameter. The error introduced due to the truncation of Eq.~(\ref{eq_taylor2}) up to the $n$-th order is given by Lagrangian remainder
\begin{align}
\delta\rho(t)=\frac{{\rm e}^{\theta{\cal L}t}}{(n+1)!}({\cal L}t)^{n+1}\rho(0) \text{,}
\end{align}
where $\theta\in(0,\,1)$. Now, we introduce a norm for the density matrix by
\begin{align}
||\rho||=\sqrt{\sum_{i,j}|\rho_{ij}|^2}=\sqrt{{\rm Tr}[\rho^{\dagger}\rho]} \text{,}
\end{align}
which is a real number that quantifies the deviation of $\rho$ from the null matrix. Depending on the perspective, this norm is mathematically called Frobenius norm if $\rho$ is treated as a matrix or spectral norm, also known as 2-norm, if $\rho$ transforms into a vector~\cite{Higham_2002}. Then, it can be shown that the norm of the Lagrangian remainder $\delta\rho(t)$ is bounded from above:
\begin{align}
||\delta\rho(t)|| & =\left|\left|\frac{{\rm e}^{\theta{\cal L}t}}{(n+1)!}({\cal L}t)^{n+1}\rho(0)\right|\right| \nonumber \\
& \le \frac{{\rm e}^{||{\cal L}||t}}{(n+1)!}(||{\cal L}||t)^{n+1}||\rho(0)|| \text{,} \label{eq_err-1}
\end{align}
where the Lindbladian norm is defined by
\begin{align}
||{\cal L}||=\sup_{M\neq 0}\frac{||{\cal L}M||}{||M||} \hspace{0.5cm}\text{for}\hspace{0.5cm} M\in\mathbb{C}^{d\times d} \text{,}
\end{align}
and the some norm properties are used~\footnote{Here, $||{\cal L}||$ is the spectral norm if ${\cal L}$ is represented as a large matrix. It is equal to the largest singular value of ${\cal L}$. The evaluation of this value requires the matrix representation of ${\cal L}$, which is computationally expensive. Alternatively, we can generate many instances of the matrix $M$ with random entries, and then estimate $||{\cal L}||$ according to its definition. The norm properties that are used to derive the upper bound are as follows. Let $A,B\in\mathbb{C}^{d\times d}$ and $\alpha\in\mathbb{C}$; we have absolute scalability $||\alpha A||=|\alpha|\cdot||A||$, sub-multiplicativity $||AB||\le||A||\cdot||B||$, and $\left|\left|{\rm e}^A\right|\right|\le{\rm e}^{||A||}$ for matrix norms.}. That upper bound can be viewed as the absolute error due to the finite-term truncation. Moreover, the relative error can also be introduced,
\begin{align}
\epsilon_r\equiv\frac{||\delta\rho(t)||}{||\rho(0)||}\le\frac{{\rm e}^{||{\cal L}||t}}{(n+1)!}(||{\cal L}||t)^{n+1} \text{.} \label{eq_err0}
\end{align}
As shown in Figure~\ref{fig_error}, the behavior of the relative error can be approximated as an exponentially decaying function of the number of terms. This allows us to achieve high accuracy with a very limited number of terms. In addition, we can also adaptively change the time or the number of retained terms in each step to gain more control over the error behavior. On the other hand, the relatively smaller number of computations in Eq.~(\ref{eq_taylor2}) generally leads to smaller accumulated round-off error. Although it is widely considered that Pad\'e approximants give more accurate results for matrix exponential, in the specific case of integrating Lindblad master equations, the proposed method is at least not worse than the vectorization method in terms of numerical accuracy. Moreover, it is necessary to point out that, although the Taylor series in the proposed method is truncated, the error thus introduced is unavoidable. As a matter of fact, the evaluation of matrix exponential based on Pad\'e approximants has the same issue. Thus, it is safe to assert that the proposed method is numerically exact. Hereafter, the proposed method is dubbed as the Taylor series method. Specifically, if terms in the Taylor series up to the ${\cal N}$-th order are retained, then the method is abbreviated as TaylorSeries${\cal N}$. The pseudo code for the Taylor series method is provided in Algorithm~\ref{algorithm}.

\begin{figure}
\centering
\begin{minipage}[t]{1.0\hsize}
\resizebox{1.0\hsize}{!}{\includegraphics{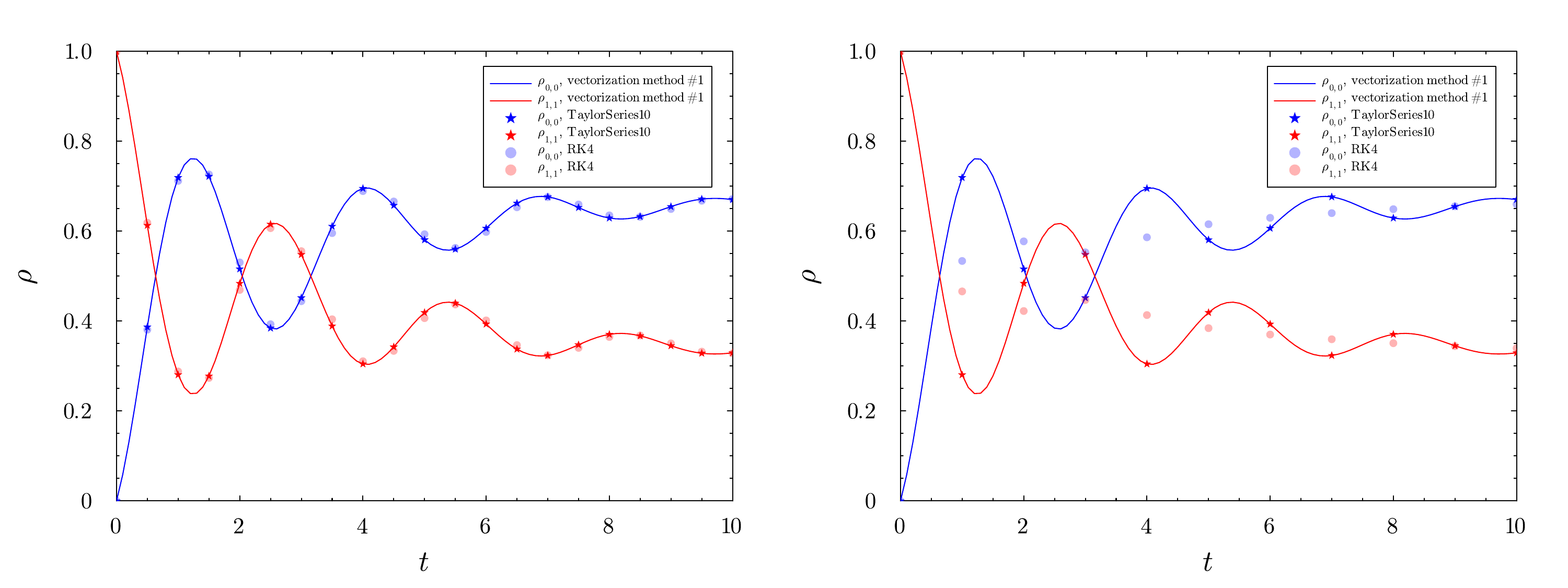}}
\end{minipage}
\caption{The time evolution of the density matrix of the dissipative two-level system initially in the excited state $\ket{1}$. The two diagonal elements are shown here. The solid lines are results calculated according to the vectorization method \#1, the starred points according to TaylorSeries10, and the filled circles according to RK4. The parameter values adopted in numerical simulations are $E=\Omega=\hbar=1.0$ and $\Gamma=0.5$. In both panels, the time step $\delta t=0.1$ is taken for the vectorization method \#1. For the other two methods, the time step $\delta t=0.5$ is taken in the left panel, and $\delta t=1.0$ in the right panel.}
\label{fig_RhoA}
\end{figure}

\section{Illustrative Examples}\label{Examples}

In order to demonstrate the validity of the Taylor series method, we consider two illustrative examples. The first one is the scenario for a one-body system whose density matrix is represented as a matrix of relatively small dimension. The Lindblad master equation is integrated with both the Taylor series method and the vectorization method. The second one is the scenario for a many-body system. In this case, all calculations are performed in tensor-network formulations. The evolution of the density matrix is calculated with the Taylor series method and also from the unraveled quantum trajectories.

\begin{figure}
\centering
\begin{minipage}[t]{0.9\hsize}
\resizebox{1.0\hsize}{!}{\includegraphics{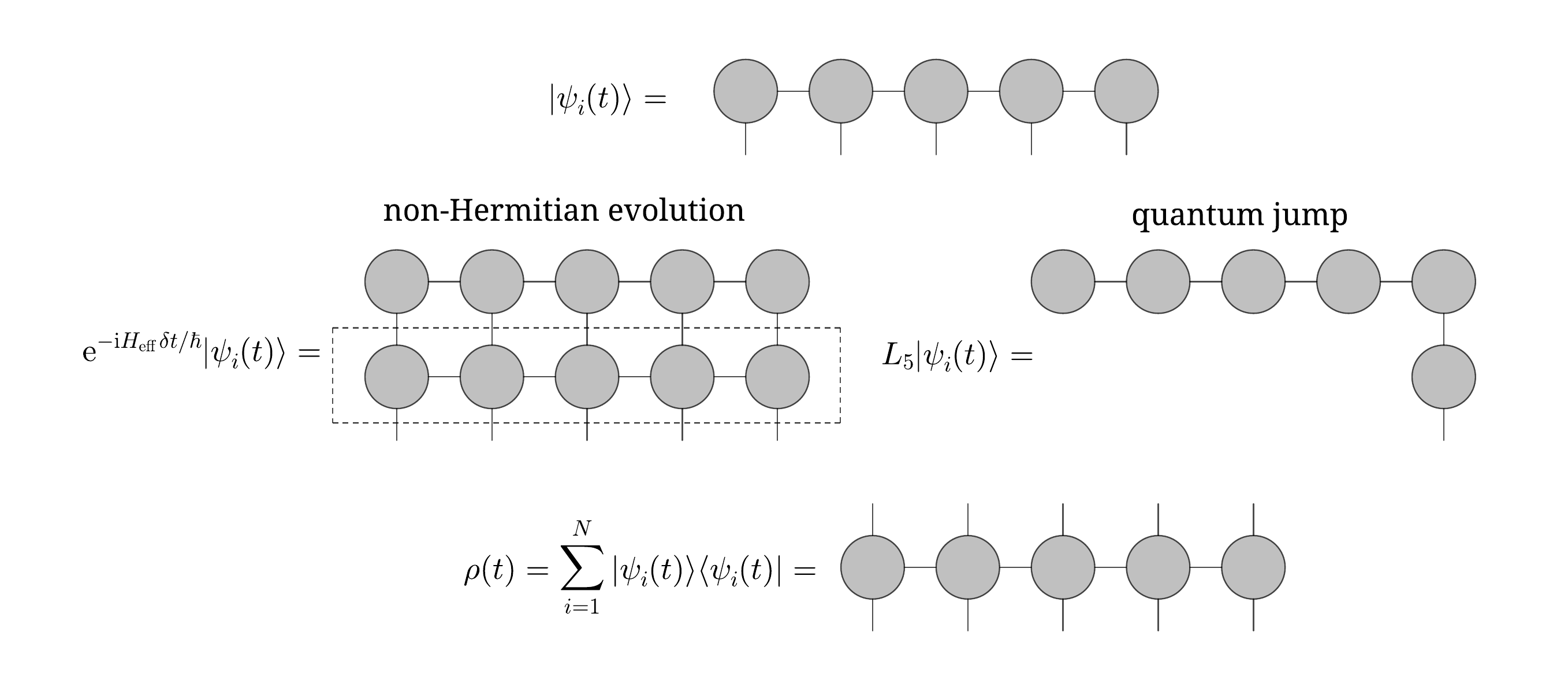}}
\end{minipage}
\caption{Quantum trajectories of the matrix product state (MPS). The ensemble-averaged evolution of the system's density matrix can be unraveled in terms of individual quantum trajectories, each consisting of abrupt jumps occurring at random times and non-Hermitian evolution in between them. The effective non-Hermitian Hamiltonian is defined by $H_{\rm eff}=H-\frac{{\rm i}\hbar}{2}\sum_iL_i^{\dagger}L_i$. The effective non-Hermitian evolution operator ${\rm e}^{-{\rm i}H_{\rm eff}\delta t/\hbar}$ is represented as a matrix product operator (MPO; circumscribed by the dashed line) that is constructed from truncated Taylor series, in the same spirit as the Taylor series method. The quantum jump as sketched in the figure corresponds to a spin lowering operator acting on the rightmost site. The density matrix is represented by matrix product density operator (MPDO, or simply MPO), which can be constructed from individual quantum trajectories.}
\label{fig_trajectory}
\end{figure}

\subsection{Two-Level System}

\par In the first example, a dissipative two-level system interacting with an electromagnetic field is considered. The Hamiltonian is given by
\begin{align}
H=E\ket{1}\!\bra{1}+\Omega\left(\ket{0}\!\bra{1}+\ket{1}\!\bra{0}\right) \text{,}
\end{align}
where we have fixed the energy of the ground state $\ket{0}$ as zero and $E$ is the energy of the excited state $\ket{1}$. Here, $\Omega$ is the frequency of driving induced by the electromagnetic field. The system then coherently switches between both states, and the populations present Rabi oscillations. This system also decays from the excited state $\ket{1}$ to the ground state $\ket{0}$ by spontaneously emitting a photon. The relevant Lindblad operator is
\begin{align}
L=\sqrt{\Gamma}\ket{0}\!\bra{1} \text{,}
\end{align}
where $\Gamma$ is the decay rate due to the coupling with the environment. For this system, we calculate the evolution of the density matrix according to the Lindblad master equation using the vectorization method \#1, the TaylorSeries10 method, and the RK4 method. The results are shown in Figure~\ref{fig_RhoA}. Initially, the system is in the pure state $\ket{1}$, i.e., $\rho_{1,1}(0)=\braket{1|\rho(0)|1}=1.0$. As time goes, the component $\rho_{0,0}=\braket{0|\rho|0}$ emerges while the trace is preserved, $\rho_{0,0}+\rho_{1,1}=1.0$. The striking agreement between the results from TaylorSeries10 and the vectorization method \#1 in both panels strongly supports the validity of the Taylor series method. In the left panel of Figure~\ref{fig_RhoA}, the time step for the RK4 is $\delta t=0.5$ , and in this case, the validity of the RK4 is expected. However, in the right panel of Figure~\ref{fig_RhoA}, where the time step for both the TaylorSeries10 and the RK4 is $\delta t=1.0$, the accuracy of RK4 is not as high as TaylorSeries10. The reason for this is obvious. In each step, the error of TaylorSeries10 is on the order ${\cal O}(\delta t^{11})$ while the error of RK4 is only on the order ${\cal O}(\delta t^{5})$. One might argue that TaylorSeries10 is more computationally expensive. It is indeed true. We will revisit this issue in Sec.~\ref{PerformanceTests}, where a fair comparison is presented. Here, we can draw a conclusion that the Taylor series method is more versatile. Because we can easily change the number of retained terms in a computer program without numerically implementing a new version. When many terms are retained, the Taylor series method becomes robust against the time step. In contrast, we need to rewrite the computer program when implementing a Runge-Kutta method of different orders.

\begin{figure}
\centering
\begin{minipage}[t]{0.6\hsize}
\resizebox{1.0\hsize}{!}{\includegraphics{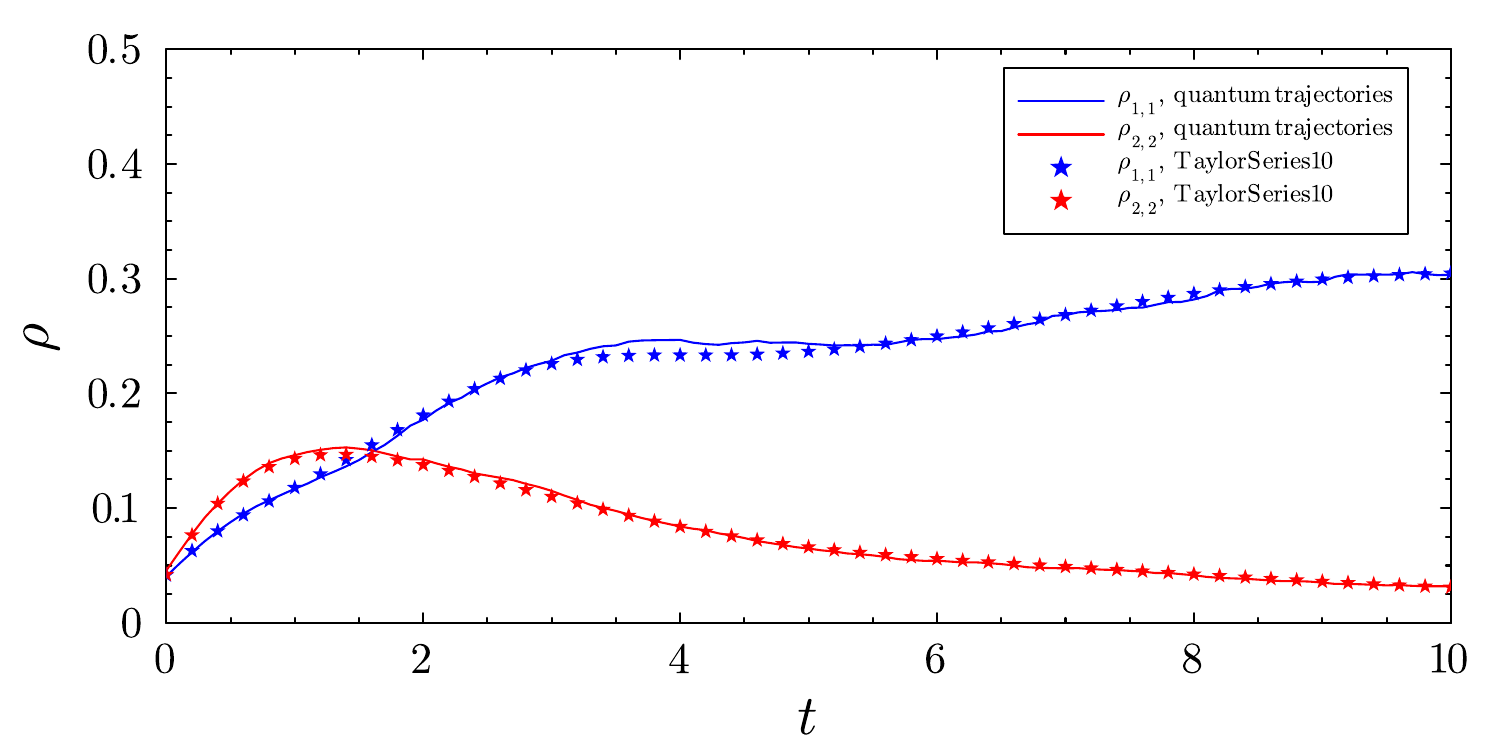}}
\end{minipage}
\caption{The time evolution of the density matrix of the dissipative Heisenberg chain, which consists of $5$ spins. The system starts its evolution from a thermal state $\rho_{\rm eq}(0)={\rm e}^{-\beta H}/{\rm Tr}[{\rm e}^{-\beta H}]$. Two diagonal elements $\rho_{1,1}\equiv\braket{\uparrow\uparrow\uparrow\downarrow\downarrow|\rho|\uparrow\uparrow\uparrow\downarrow\downarrow}$ and $\rho_{2,2}\equiv\braket{\uparrow\downarrow\downarrow\downarrow\downarrow|\rho|\uparrow\downarrow\downarrow\downarrow\downarrow}$ are shown as representives. The solid lines are results calculated from quantum trajectories, while the starred points are according to the TaylorSeries10. The parameter values $J=\Gamma=\hbar=\beta=1.0$ are adopted in numerical simulations. In the quantum jump method, the time step $\delta t=0.1$ is taken, and $N=1000$ stochastic trajectories are generated to give the average evolution behavior. In TaylorSeries10, the time step $\delta t=0.2$ is taken. The diagonal elements are calculated by first representing pure states (ket and bra) as matrix product states (MPSs), and then contracting with MPO representation of the density matrix from both the bottom and the top sides. In the tensor-network calculations, the default parameter value \texttt{cutoff=1.0e-14} is used when an MPO is applied to an MPO or MPS as the singular value decompositions (SVDs) are performed. This default parameter is defined as a real number $\epsilon$ by $\sum_{n\in{\rm discarded}}\sigma_n^2/\sum_n\sigma_n^2<\epsilon$, where $\{\sigma_n\}$ are singular values. The minor disagreement between the results from the two methods is due to the fluctuations in stochastic quantum trajectories.}
\label{fig_RhoB}
\end{figure}

\subsection{Heisenberg Chain}

\par The second example is a boundary-driven dissipative Heisenberg chain with the Hamiltonian given by
\begin{align}
H & =-J\sum_{i=1}^{L-1}{\bf S}_i\cdot{\bf S}_{i+1} \nonumber \\
& =-J\sum_{i=1}^{L-1}\left(S_i^xS_{i+1}^x+S_i^yS_{i+1}^y+S_i^zS_{i+1}^z\right) \text{,}
\end{align}
where ${\bf S}_i=\left(S_i^x,S_i^y,S_i^z\right)$ is the spin operator at the $i$-th site defined as half the Pauli matrices, and $J$ the coupling constant. Through a Jordan-Wigner transformation, this system can be mapped into a spinless fermion Hubbard model with nearest-neighboring interactions. It models dc transport of fermions in a quantum wire if the two edges are in contact with reservoirs, emboding Markov channels. As such, two Lindblad operators
\begin{align}
L_1=\sqrt{2\Gamma}S_1^+ \text{,}\hspace{0.5cm} L_L=\sqrt{2\Gamma}S_L^- \text{,}
\end{align}
are introduced at the leftmost site (source) and rightmost site (drain), respectively. Here, $S_1^+\equiv S_1^x+{\rm i}S_1^y$ and $S_L^-\equiv S_L^x-{\rm i}S_L^y$ are the spin raising and lowering operators.

\par In the following, we exploit the power of tensor networks to perform numerical calculations. The density matrix is represented as a matrix product density operator (MPDO) or simply matrix product operator (MPO)~\cite{Verstraete_PhysRevLett_2004b}, as schematically illustrated in Figure~\ref{fig_trajectory}. The Hamiltonian and Lindblad operators are also represented as MPOs~\cite{Pirvu_NewJPhys_2010}. The Taylor series method works very well in tensor-network formulations; the matrix multiplications on $\rho(t)$ from the left and right sides are now converted into the MPO contractions on $\rho(t)$ from the bottom and top sides. For generality, the system is initially prepared in a thermal equilibrium state. Then, we use the Taylor series method to calculate the density matrix at subsequent times. Some elementary details of the tensor-network implementation of the Taylor series method can be found in Appendix~\ref{TN}.

\par The vectorization method can also be employed to study the mixed-state dynamics of one-dimensional lattice systems with tensor networks~\cite{Zwolak_PhysRevLett_2004}. The basic idea is to turn the MPDO into the matrix product state (MPS). In the language of tensor-network diagrams, it can be regarded as reshaping one of the legs and gluing it to the other. Then, the time-evolving block decimation (TEBD)~\cite{Vidal_PhysRevLett_2003, Vidal_PhysRevLett_2004} or the time-dependent variational principle (TDVP)~\cite{Haegeman_PhysRevLett_2011, Haegeman_PhysRevB_2016} can be used to simulate the real-time Markovian dynamics if the Lindbladian decomposes into terms of nearest-neighbor couplings. In this way of simulation, however, the positivity of the density matrix is not guaranteed, and checking the positivity is known to be an issue~\cite{Kliesch_PhysRevLett_2014}. A possible workaround is the algorithm provided in Ref.~\cite{Werner_PhysRevLett_2016}, where at every stage $\rho$ is kept in its locally purified $\rho=XX^{\dagger}$, thereby ensuring positivity at all times during the evolution. In order to simplify the numerical implementation, we choose to benchmark the Taylor series method with the quantum jump method, which makes use of a stochastic unraveling of the master equation and then employs pure state techniques. Moreover, this method can work along with the tensor networks, i.e., generating quantum trajectories of pure states represented as MPSs. In the numerical simulation, $L=5$ spins are specified for the system, and $N=1000$ quantum trajectories $\{\ket{\psi_i(t)}\}_{i=1}^N$ are sampled starting from a thermal equilibrium state $\rho_{\rm eq}(0)={\rm e}^{-\beta H}/Z$, where $Z\equiv{\rm Tr}[{\rm e}^{-\beta H}]$ is the partition function. Because the stochastic evolution of quantum trajectories starts from pure states, this initial thermal equilibrium state is unraveled into a set of typical states $\{\ket{\psi_i(0)}\}_{i=1}^N$ using the algorithm called minimally entangled typical thermal states (METTS)~\cite{White_PhysRevLett_2009, Stoudenmire_NewJPhys_2010}. See Appendix~\ref{METTS} for a detailed account of this algorithm. The density matrix at subsequent times can be constructed as follows
\begin{align}
\rho(t)\approx\frac{1}{N}\sum_{i=1}^N\ket{\psi_i(t)}\!\bra{\psi_i(t)} \text{.} \label{eq_rhot}
\end{align}
The density matrix constructed in this way is positive and trace-preserving. Since the trajectories are independent from each other, the statistical error associated with the constructed density matrix is estimated as $\Delta\rho\sim 1/\sqrt{N}$, which is down to a few percent if $N$ is more than one thousand. Figure~\ref{fig_trajectory} shows the generality of the quantum jump method. The effective non-Hermitian evolution ${\rm e}^{-{\rm i}H_{\rm eff}\delta t/\hbar}$ is calculated with the Taylor series method, where the effective non-Hermitian Hamiltonian is defined by $H_{\rm eff}=H-\frac{{\rm i}\hbar}{2}\sum_iL_i^{\dagger}L_i$. This evolution can also be realized with TEBD or TDVP, which actually gives the same result as that of the Taylor series method. This indicates that the Taylor series method can also be used to simulate the dynamics of pure states. In Figure~\ref{fig_RhoB}, two diagonal elements of the density matrix calculated from quantum trajectories are compared with the results obtained from the Taylor series method. The overall agreement is found, and this again supports the validity of the Taylor series method. In the numerical simulation, the small number of sites ($L=5$) is specified for the Heisenberg chain. The purpose of this deliberate choice is that the diagonal elements of density matrix still have a significant fraction of ${\rm Tr}[\rho(t)]=1$ so that the noise from the quantum trajectories is relatively small. The Taylor series method integrated with tensor networks can be easily applied to study much longer spin chains. The computer program for two illustrative examples is coded in \textsc{julia}, and the ITensor library~\cite{Fishman_SciPostPhysCodeb_2022} is used additionally for the tensor-network calculations in simulating the dissipative dynamics of the Heisenberg chain. Interested readers are also referred to Ref.~\cite{Psarras_arXiv_2022} for a comprehensive and up-to-date snapshot of software for tensor computations.

\section{Performance Tests}\label{PerformanceTests}

Detailed performance tests of the Taylor series method are now carried out to show its advantages over the vectorization method and the RK4 method. The dissipative Heisenberg chain is used as the benchmarking system, as the underlying Hilbert space grows exponentially with the number of sites. The density matrix $\rho$, Hamiltonian $H$, and Lindblad operators $\{L_1,L_L\}$ are all represented as matrices, without usage of tensor networks.

\subsection{Advantage over the Vectorization Method}

\begin{figure}
\centering
\begin{minipage}[t]{0.49\hsize}
\resizebox{1.0\hsize}{!}{\includegraphics{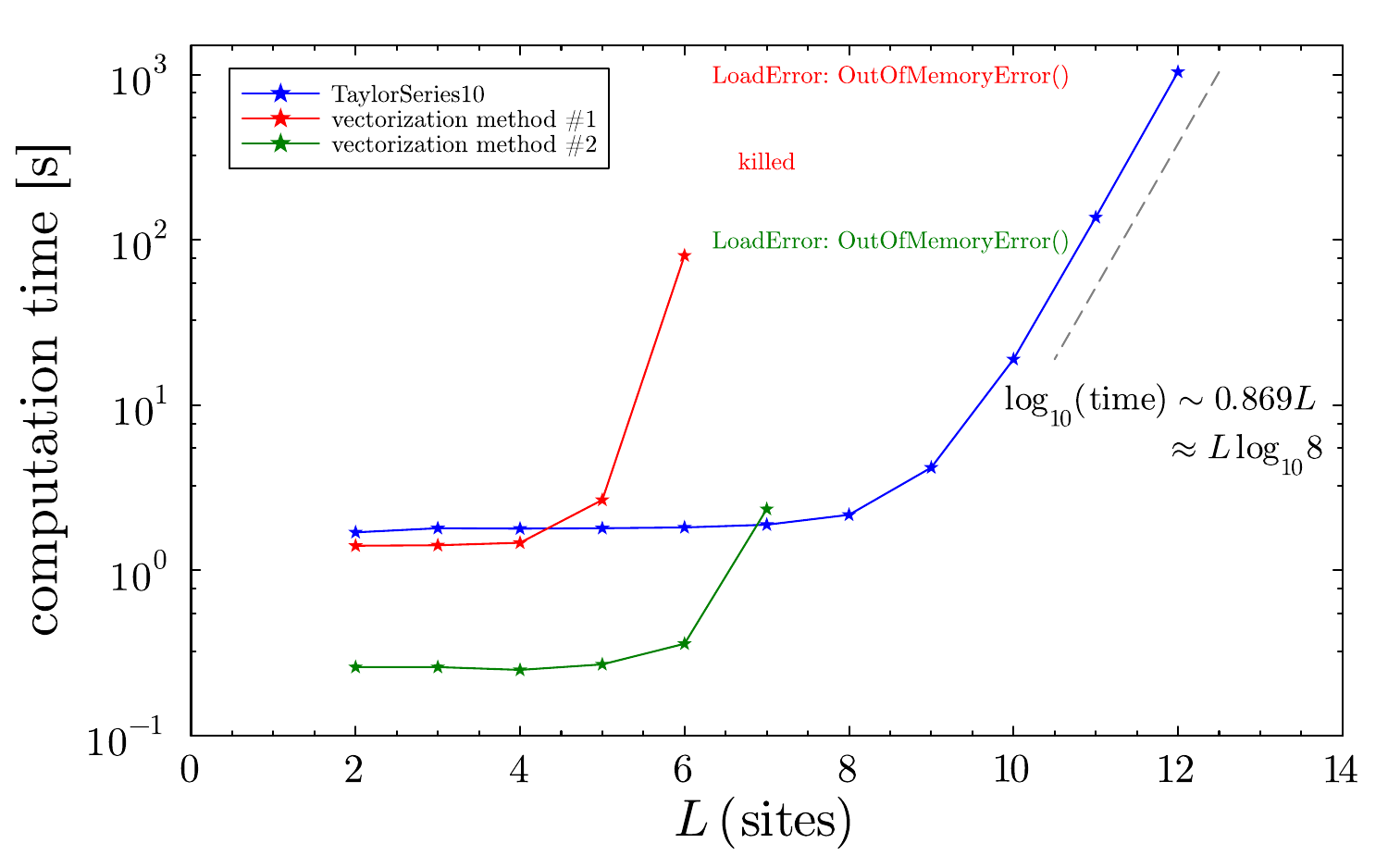}}
\end{minipage}
\begin{minipage}[t]{0.49\hsize}
\resizebox{1.0\hsize}{!}{\includegraphics{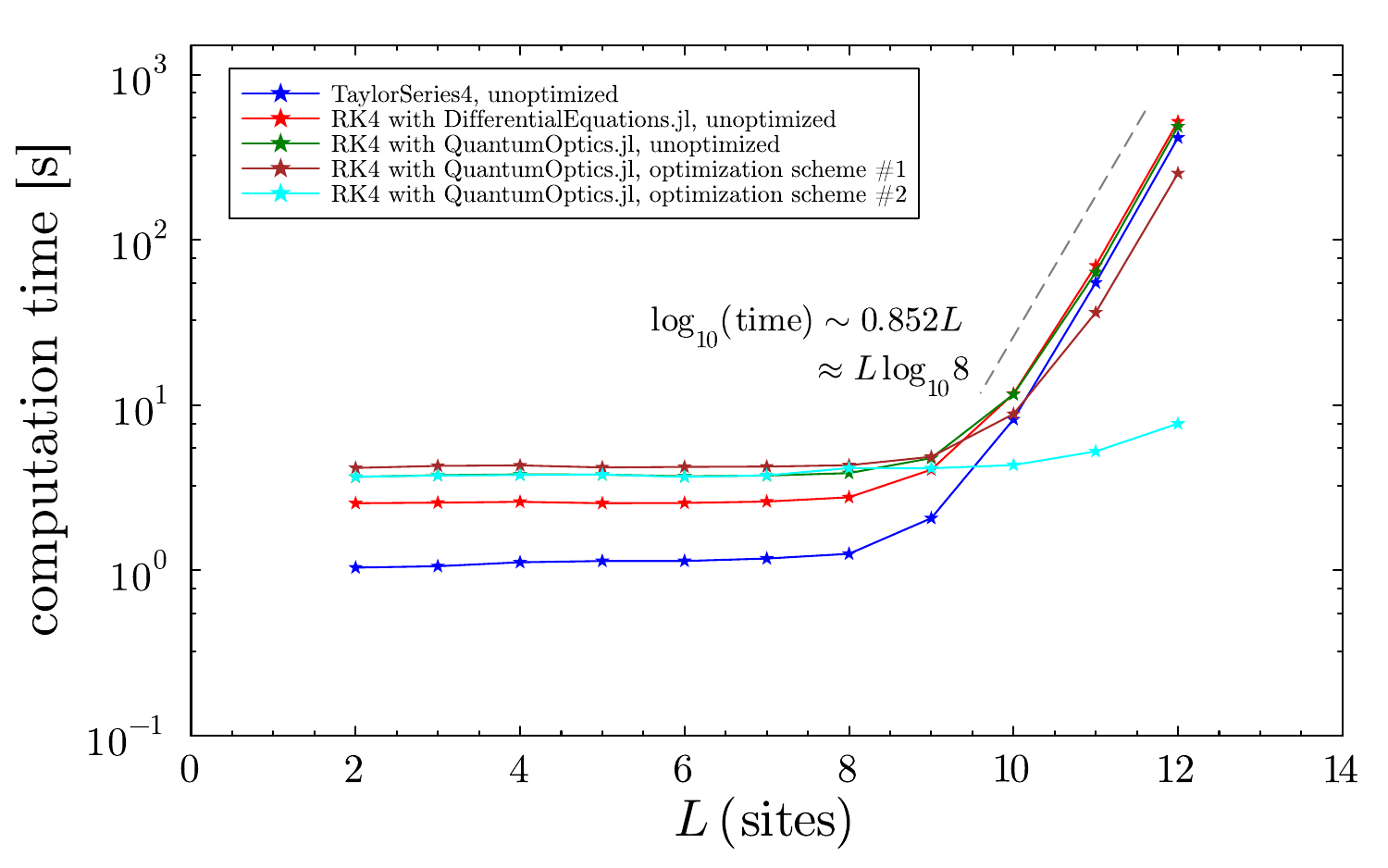}}
\end{minipage}
\caption{Benchmarks measuring the time elapsed when performing time evolution of the dissipative Heisenberg chain in one step according to the Lindblad master equation. All operators are represented as matrices. The computer program is coded in \textsc{julia} in single-thread mode. The computing platform is a PC with an Intel Core i7-12700H processor, Ubuntu 24.04.1 LTS OS, and 16 GB RAM. In the left panel, comparisons are made between TaylorSeries10 and the vectorization method, both \#1 and \#2. For vectorization method \#1, the largest number of sites of the system that can be handled by the computer program is $L=6$. When $L=7$, the computer program was killed after long-time running, and when $L\ge 8$, the out-of-memory error occurred due to insufficient memory to store the Lindbladian ${\cal L}$ in the matrix representation. For vectorization method \#2, the out-of-memory error also occurred due to the same reason when $L\ge 8$. In vectorization method \#1, the built-in function \texttt{exp} is used for matrix exponential. As in TaylorSeries10, terms up to the 10-th order are retained in vectorization method \#2. In the right panel, comparisons between TaylorSeries4 and RK4 are made and, as shown in the legend, the latter is implemented in different ways. Two optimization techniques are identified: one adopting the effective non-Hermitian Hamiltonian and the other using sparse matrices. In optimization scheme \#1, only the first technique is applied. In optimization scheme \#2, both techniques are applied. In both panels, the scaling behavior of computation time with the number of sites is observed, $\log_{10}({\rm time})\sim L\log_{10}8$, as expected.}
\label{fig_test}
\end{figure}

The advantage of the Taylor series method over the vectorization method is obvious from the comparison in numerical complexities presented in Table~\ref{tab_comparison}. To show an intuitive picture of this advantage, we make some numerical calculations and schemitize the results in the left panel of Figure~\ref{fig_test}. In this figure, the computation time for one integration step is plotted against the number of sites. The clear message from this figure is that the Taylor series method is indeed more numerical efficient, as expected. For the vectorization method, the first barrier to encounter is the insufficient memory to store the Lindbladian ${\cal L}$ in matrix representation. In the practical numerical calculation, the maximum number of sites that the vectorization method \#1 can handle with 16 GB RAM (Random Access Memory) is $6$. In this case, the dimension of the underlying Hilbert space is $d=2^6=64$, and the Lindbladian ${\cal L}$ is a $4096\times 4096$ matrix, with each element a complex number occupying 16 bytes. Simple arithmetic calculation gives 268435456 bytes (equivalently 0.25 GB) required for the storage of such Lindbladian matrix. When the dissipative Heisenberg chain takes $L=7$ sites, the required storage for ${\cal L}$ becomes 4 GB, which is so demanding. In the practical numerical calculation with vectorization method \#1, the program was killed after running for a long time. However, the numerical calculation with vectorization method \#2 still gave the result due to the lower complexity in time. When $L\ge 8$, the storage of a single ${\cal L}$ requires at least 64 GB. As a consequence, \texttt{LoadError: OutOfMemoryError()} immediately prompted after executing the program coded with the vectorization method, both \#1 and \#2. In the Taylor series method, the numerical computation scales with ${\cal O}(d^3)$ with $d=2^L$. So, we have the scaling relation for the computation time, $\log_{10}({\rm time})\sim L\log_{10}8$, as indeed observed in the left panel of Figure~\ref{fig_test}. The plateau of the computation time for a small number of sites finds its origin from the underlying mechanism of matrix multiplications. The issue of large memory consumption also exists in the Taylor series method. From the above analysis, we judge that the maximum number of sites that the Taylor series method can handle is twice as large. This issue can be alleviated with the usage of tensor networks.

\subsection{Advantage over the RK4 Method}

In the previous text, we have pointed out that the accurary of the Runge-Kutta methods is time-sensitive. When implementing the Runge-Kutta method of different orders, we should rewrite the code. What makes the thing more complicated is that the coefficients of Runge-Kutta methods of higher orders are not uniquely determined. In contrast, the Taylor series method is more versatile as the number of retained terms can be easily changed without the need to implement a new version. As such, the accuracy of the Taylor series method is more robust against the time step if more terms are kept. However, in order to guarantee high accuracy, the strategy of the Runge-Kutta methods is to take very small integration steps. In this way, it demands more computational cost. This issue is the same as the Taylor series method in which more terms are retained to guarantee the accuracy. In order to draw a fair comparison between the Taylor series method and the Runge-Kutta methods, comprehensive evaluations of their performance are needed. One additional remark on the application of the Runge-Kutta methods to solve Lindblad master equations is in order here. Runge-Kutta methods are applicable generally to cases of linear and nonlinear ordinary differential equations. However, Lindblad master equations are linear and, in this case, there exists a specific and more accurate solver that is given by the action of Lindbladian exponential on the initial density matrix.

The classic RK4 is here compared with the Taylor series method. As shown in Appendix~\ref{RK4}, it is simple to write code for RK4 from scratch. However, we choose to call relevant solvers from popular numerical packages so that we can gain insight into their performance. Since we write code primarily in \textsc{julia}, two packages in the same programming language are used: QuantumOptics.jl~\cite{Kramer_ComputPhysCommun_2018} and DifferentialEquations.jl~\cite{Rackauckas_JOpenResSoftw_2017}. The former is a numerical framework that makes it easy to simulate various kinds of open quantum systems, while the latter is a suite for numerically solving differential equations. It should be noted that in QuantumOptics.jl various quantum systems are defined through quantum objects, such as states and operators, and then their evolutions are simulated by calling underlying solvers from DifferentialEquations.jl. In other words, the solvers in QuantumOptics.jl are basically the wrappers of the solvers from DifferentialEquations.jl.

The benchmarks for performance comparisons between the self-implemented TaylorSeries4 and the RK4 solvers from both packages are presented in the right panel of Figure~\ref{fig_test}. In this figure, the computation times for the one-step evolution according to the Lindblad master equation are shown in circumstances where the Heisenberg chain takes a different number of sites and the system is simulated in different ways. The system is simulated with the RK4 solver from QuantumOptics.jl and also more directly with the RK4 solver from DifferentialEquations.jl. Through the comparison in this way, we get to know whether and how the RK4 solver in QuamtumOptics.jl optimizes the system before calling the underlying RK4 solver from DifferentialEquations.jl. As a consequence, we did spot two optimization techniques that the solvers in QuantumOptis.jl use. The first one is to use the effective non-Hermitian Hamiltonian. Specifically, the solvers in QuantumOptics.jl first define the effective non-Hermitian Hamiltonian $H_{\rm eff}=H-\frac{{\rm i}\hbar}{2}\sum_iL_i^{\dagger}L_i$, and then simulate the system according to the following Lindblad master equation
\begin{align}
\frac{{\rm d}\rho}{{\rm d}t}=-\frac{\rm i}{\hbar}\left[H_{\rm eff},\rho\right]+\sum_iL_i\rho L_i^{\dagger} \text{.} \label{eq_Lindblad2}
\end{align}
For quantum systems with large Hilbert space, the most critical parts of computation are matrix multiplications. Compared with the original Lindblad master equation~(\ref{eq_Lindblad}), the number of matrix multiplications in Eq.~(\ref{eq_Lindblad2}) is indeed reduced. The other optimization technique is the usage of sparse matrices that lead to considerable speed-ups. It should be pointed out that these two optimization techniques are not specific to QuantumOptics.jl; they can also be used in the Taylor series method. For two reasons: (1) considering general circumstances and (2) evaluating the computation time in terms of matrix multiplications, we compare the performance of the unoptimized TaylorSeries4 and the unoptimized RK4 solvers from both packages~\footnote{In \textsc{julia}, the built-in vectors and matrices are by default dense. The vectors/matrices for the solvers from DifferentialEquations.jl are also dense by default. In QuantumOptics.jl, the default master equation solver \texttt{timeevolution.master()} internally first creates the effective non-Hermitian Hamiltonian and then simulates the system with the corresponding Lindblad master equation. If for any reason this behavior is unwanted, the solver \texttt{timeevolution.master\_h()} (indicating that Hermitian Hamiltonian is internally used) is available for use. Moreover, the internally constructed sparse operators can be converted to dense operators by the function \texttt{DenseOperator()}.}. In this case, we notice approximately the same computation times needed for these solvers, as shown by the blue, red, and green lines in the right panel of Figure~\ref{fig_test}. It should be noted here that TaylorSeries4 and RK4 have the same numerical cost as predicted by the number of matrix multiplications. As expected, the scaling behavior $\log_{10}({\rm time})\sim L\log_{10}8$ is also observed here. The clear message from this comparison is that, if the optimization techniques are not applied, the RK4 solvers from the two packages show no advantage over the self-implemented TaylorSeries4. The number of matrix multiplications serves as a reliable measure for the computation cost. In the right panel of Figure~\ref{fig_test}, we also present the cases where optimization techniques are applied for the RK4 solver in QuantumOptics.jl. It is apparent that considerable speed-ups are achieved, especially when sparse matrices are used. Indeed, most entries in the matrices representing the Hamiltonian and the Lindblad operators for the dissipative Heisenberg chain are zero.

\begin{figure}
\centering
\begin{minipage}[t]{1.0\hsize}
\resizebox{1.0\hsize}{!}{\includegraphics{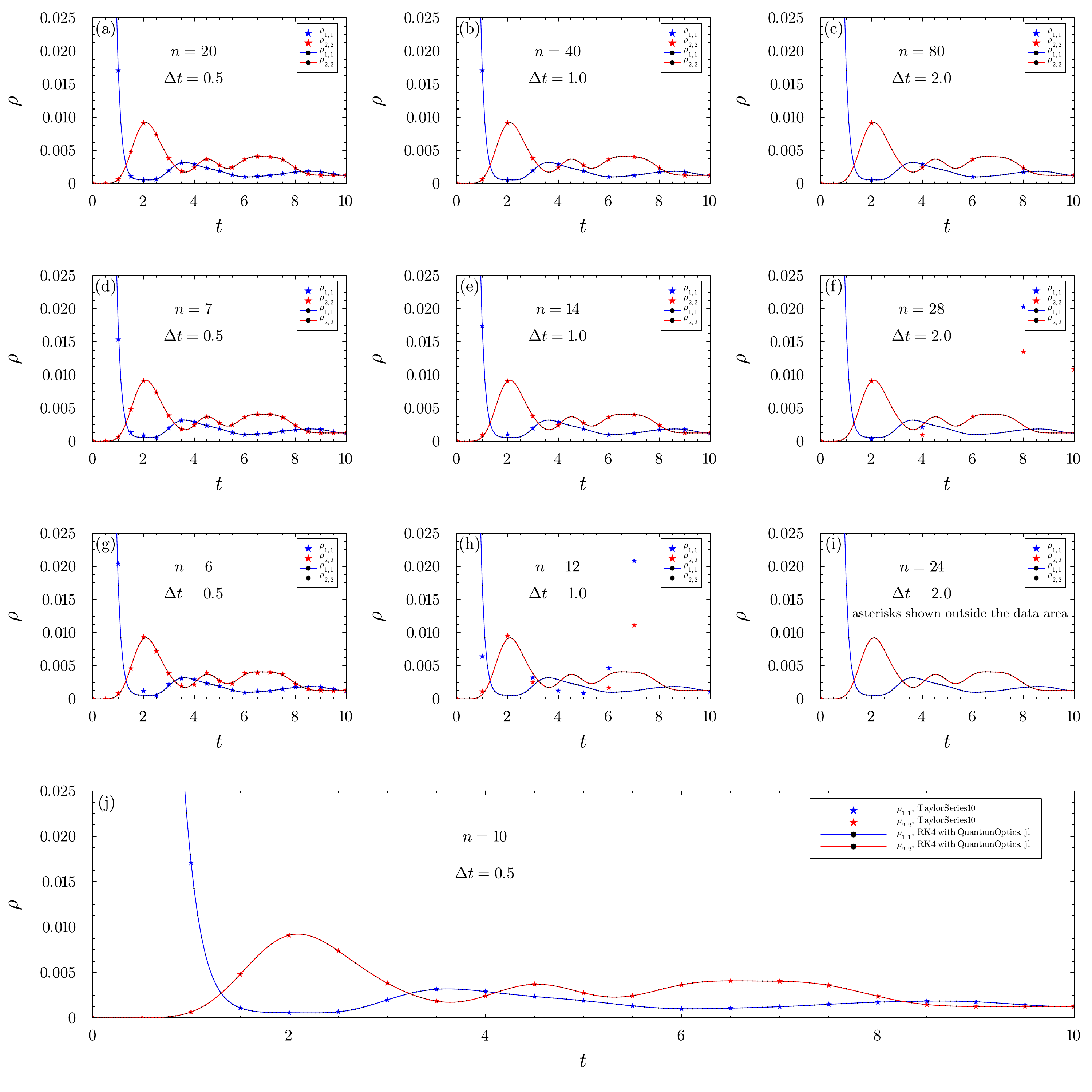}}
\end{minipage}
\caption{The time evolution of the density matrix of the dissipative Heisenberg chain consisting of 9 spins. The system is initially in the pure state $\ket{\psi_1}=\ket{\uparrow\downarrow\uparrow\downarrow\uparrow\downarrow\uparrow\downarrow\uparrow}$. Two diagonal elements $\rho_{1,1}\equiv\braket{\psi_1|\rho|\psi_1}$ and $\rho_{2,2}\equiv\braket{\psi_2|\rho|\psi_2}$ are shown, where $\ket{\psi_2}\equiv\ket{\downarrow\uparrow\downarrow\uparrow\downarrow\uparrow\downarrow\uparrow\downarrow}$. The parameter values for the system are $J=\Gamma=\hbar=1$. In all panels, the solid lines joining the black dots represent the solutions from calling an unoptimized QuantumOptics.jl solver that is specified to use RK4 as the underlying method. The asterisks are solutions solved with the Taylor series method. For panels from (a)-(i), the QuantumOptics.jl solver is specified to have fixed integration time step $\delta t=0.1$, separating every two neighboring black dots. The sampling time step $\Delta t$ and the number of retained terms $n$ in the Taylor series method take different values, as shown in each panel. In the panels in each row, the computational costs for the Taylor series method are the same, as they are proportional to $n/\Delta t$. In the last panel (j), the sampling time step is $\Delta t=0.5$ and TaylorSeries10 is used. In this panel, the QuantumOptics.jl solver is specified to take adaptive integration time steps, and thus the black dots representing the internal solutions are unevenly distributed along the time axis. The computation times are 51 seconds for TaylorSeries10 and 271 seconds for the QuantumOptics.jl solver. The computing platform is a PC with an Intel Core i7-12700H processor, Ubuntu 24.04.1 LTS OS, and 16 GB RAM.}
\label{fig_RhoC}
\end{figure}

The strategy for RK4 to guarantee the accuracy is to integrate differential equations with small time steps. In this scenario, two kinds of time steps should be differentiated: the time step for internal integration and the time step between sampling points for output. From now on, we use $\delta t$ to denote the former and $\Delta t$ to denote the latter. For RK4, the time step for internal integration is usually much smaller, $\delta t<\Delta t$. However, for the Taylor series method, it is not necessary to differentiate the two. Because the Taylor series method can guarantee the accuracy by retaining more terms. The strategies to improve the accuracy for the RK4 and the Taylor series method are both accompanied by more computational cost, i.e., the time. In the following, we make a fair comparison which one gives more accurate results with the same amount of computational cost. As already demonstrated before, the computational time is proportional to the number of matrix multiplications or, equivalently, by operations of the Lindbladian ${\cal L}$ applied on the density matrix $\rho$. In one integration step $\delta t$, RK4 has 4 such operations. However, in one integration step $\Delta t$, the Taylor series method with $n$ retained terms has $n$ such operations. Let $\Delta t=m\delta t$, then, to output one sampling point, RK4 needs to perform $4m$ such operations. The condition for the same amount of computational cost requires that $4m=n$. The error of RK4 to output one sampling point is on the order ${\cal O}(\delta t^4)$, whereas for the Taylor series method with $n$ retained terms it is on the order ${\cal O}(\Delta t^{n+1})$. We hope that the Taylor series method is more accurate, so we expect the inequality
\begin{align}
\Delta t^{n+1}<\delta t^4 \label{eq_err1}
\end{align}
to hold. Substituting $\Delta t=m\delta t$ and $n=4m$ into this inequality yields
\begin{align}
\delta t<m^{-\frac{4m+1}{4m-3}}=m^{-1-\frac{4}{4m-3}} \text{.}
\end{align}
Considering that $m$ is typically on the order ${\cal O}(10)$, the above relation approximately reduces to
\begin{align}
m\delta t<1 \text{.} \label{eq_err2}
\end{align}
This suggests that when $\Delta t=m\delta t<1$, the inequality~(\ref{eq_err1}) indeed holds and therefore the Taylor series method is more accurate than RK4 in this case. For the inequality~(\ref{eq_err1}) it can be intuitively understood that the left-hand side exponentially decays with $n$ when $\Delta t<1$ while the right-hand side polynomially decays as $\delta t$ decreases; the former decays much faster. However, since the calculations are on the orders, the right-hand side of inequality~(\ref{eq_err2}) should also be intepreted as the order, that is ${\cal O}(1)$. On the other hand, the inequality~(\ref{eq_err2}) also suggests that there exists a time $T$ that the Taylor series method gives less accurate results when $\Delta t>T$. So, there exists an accuracy-to-inaccuracy transition across an undetermined time $T$.

The above prediction of the existence of the transition is checked in Figure~\ref{fig_RhoC}, where two diagonal elements of the density matrix of the Heisenberg chain consisting of $9$ sites are plotted as functions of time. In all panels, the overall computational costs for the Taylor series method are quantified by a proportional factor $n/\Delta t$. So, we can see that the Taylor series method in three panels in each row has the same overall computational costs. In panels~(a)-(i), the integration time step for RK4 takes a fixed value $\delta t=0.1$. So, $m=\Delta t/\delta t$ can be calculated with the values of $\Delta t$ that are shown. For the panels~(a)-(c) in the top row, the equality $n=4m$ holds, suggesting that the Taylor series method has the same computational cost as RK4. In these three panels corresponding to $\Delta t=0.5,1.0,2.0$ respectively, the results from the Taylor series method agree well with those from RK4. However, the accuracy-to-inaccuracy transition across an undetermined time $T$ is not observed, possibly indicating that $T>2.0$. For the panels (d)-(f) in the second row, the Taylor series method is less costly in computation time, $n<4m$, and in this case we find that the Taylor series method gives inaccurate results when $\Delta t=2.0$, as shown in the panel~(f). This implies that the transition happens at $T\approx 2.0$. In the panels~(g)-(i) in the third row, the Taylor series method consumes even less computation time, and in this case, the transition happens at $T\approx 1.0$. The existence of the transition is confirmed. Now, we are ready to give the following assertion: Given equal or less computational cost and for typical sampling time steps that are not too large, the Taylor series method can still produce results that are more accurate than or at least as accurate as those from RK4. Here, it should be pointed out that the error of the Taylor series method on the order ${\cal O}(\Delta t^{n+1})$ is actually overestimated, because there is a factor $(n+1)!$ in the denominator of Eq.~(\ref{eq_err0}). The last panel~(j) in Figure~\ref{fig_RhoC} corresponds to the case where the RK4 solver takes adaptive time steps for integration. This is the default integration scheme for the solvers in QuantumOptics.jl. In this panel, TaylorSeries10 is adopted to produce results for comparison. The sampling time step is $\Delta t=0.5$, and within two consecutive sampling points, there are approximately 10 unevenly distributed points representing the internal solutions from RK4. From this panel, striking agreement is noticed between the results from the two methods. However, according to the preceding analysis, the computation cost for TaylorSeries10 is less than that for RK4. Indeed, in practical calculations, TaylorSeries10 consumes 51 seconds whereas RK4 consumes 271 seconds. It is noted that, in this last panel, the underlying RK4 solver is called by a QuantumOptics.jl solver at a higher level. Therefore, the advantage of the Taylor series method over numerical package QuantumOptics.jl is clearly demonstrated. Note that the Taylor series method can also be implemented with adaptive time steps by controlling the absolute error~(\ref{eq_err-1}) or relative error~(\ref{eq_err0}) under a certain level. Alternatively, we feel that it is more elegant to adaptively control the number of retained terms in the Taylor series method. For time-dependent systems, the time step of the Taylor series method matters. For example, in time crystals, time steps should be such intervals within which the Hamiltonian or Lindblad operators do not change drastically. In this case, we can employ an adaptive scheme by simultaneously changing the time steps and the orders of truncation. According to the errors~(\ref{eq_err-1}) and~(\ref{eq_err0}), when a small time step is taken, the order of truncation can also be reduced. Moreover, when the long-time dynamics of open quantum systems is studied, it is helpful to use the Taylor series method with large time steps.

\section{Conclusion and Discussion}\label{Conclusion}

In this work, we have proposed a generic method for integrating Lindblad master equations. This method is simple and straightforward, exploiting the definition of the Lindbladian exponential expressed as the Taylor series expansion. The action of the Lindbladian directly transforms into the multiplications of the Hamiltonian and Lindblad operators on both sides of the density matrix. The validity of this Taylor series method has been sufficiently demonstrated with two benchmark cases, with the second one in the formulation of tensor networks. Compared with the vectorization method, the Taylor series method is apparently more numerically efficient, thus allowing one to tackle open quantum systems whose underlying Hilbert space is much larger. We also compared this Taylor series method with the fourth-order Runge-Kutta method, and revealed that the former is more versatile and is also more numerically efficient to some extent. To conclude, the Taylor series method is more advantageous, and we expect that it facilitates further research into the nonequilibrium dynamics of open quantum systems. In addition, we also hope that this Taylor series method can be quickly appreciated in the community and be implemented as one of the standard solvers in popular numerical packages.

In the end, some discussions are in order. The core idea of the Taylor series method is that the vectorization of the density matrix is not necessary at all. In this line, future developments can be explored. The Lindbladian operator is, with no doubt, sparse if represented as a large matrix for systems with many degrees of freedom. The Krylov subspace method can be potentially used to project the Lindbladian onto a relatively small Krylov subspace where calculating the exponential is computationally much cheaper. In this method, the Krylov subspace is expanded by several matrices in analogy with vectors. The action of the Lindbladian operator on matrices follows the Lindblad master equation. If the upper Hessenberg matrix resulting from the projection of the Lindbladian on the Krylov subspace is comparable in size with the density matrix, then the same numerical complexity as the Taylor series method can be achieved. However, the method is not exact unless the dimension of the Krylov subspace is equal to the dimension of the Lindbladian operator. Despite that, the potential of this method deserves to be explored. Moreover, the advanced techniques such as Magnus expansions and high-order commutator-free schemes can also be explored for application in the cases where the Lindblad master equation is time-dependent. Since the Lindbladian operator is not explicitly represented as a large matrix, how to apply these techniques is still unclear.

\section*{Acknowledgements}
We acknowledge two anonymous referees for their critical comments that have helped to extensively clarify the paper. Fan Zhang thanks H. T. Quan for his encouragement and support. This work was supported financially by the National Natural Science Foundation of China (NSFC) under Grant No. 12505048 and the JST Moonshot R\&D under Grant No. JPMJMS226B.

\appendix

\section{The Fourth-Order Runge-Kutta Method}\label{RK4}

The widely used techniques for numerically solving ordinary differential equations (ODEs) are Runge-Kutta methods, of which the fourth-order version (RK4) is the most well-known one. It offers a good balance between computational efficiency and accuracy. Let's consider an initial value problem (IVP) such as the Lindblad master equation
\begin{align}
\frac{{\rm d}\rho}{{\rm d}t}={\cal L}\rho=\frac{\rm i}{\hbar}[\rho,H]+\sum_i\left(L_i\rho L_i^{\dagger}-\frac{1}{2}\{L_i^{\dagger}L_i,\rho\}\right)
\end{align}
with the initial condition $\rho(0)=\rho_0$. Our goal is to compute $\rho(t)$ at successive points using time steps $\delta t$, such that $\rho_{n+1}=\rho(t_n+\delta t)$ is approximated based on $\rho_n=\rho(t_n)$. The RK4 method estimates $\rho_{n+1}$ using a weighted average of slopes at four points within the interval $[t_n,\,t_n+\delta t]$. Specifically, it computes four slopes:
\begin{align}
& k_1={\cal L}\left(\rho_n\right) \text{,} \\
& k_2={\cal L}\left(\rho_n+\frac{\delta t}{2}k_1\right) \text{,} \\
& k_3={\cal L}\left(\rho_n+\frac{\delta t}{2}k_2\right) \text{,} \\
& k_4={\cal L}\left(\rho_n+\delta tk_3\right) \text{,}
\end{align}
and then updates $\rho_{n+1}$ as
\begin{align}
\rho_{n+1}=\rho_n+\frac{\delta t}{6}(k_1+2k_2+2k_3+k_4) \text{.}
\end{align}
The RK4 method is fourth-order, meaning that the local integration error is on the order ${\cal O}(\delta t^5)$, while the total accumulated error is on the order ${\cal O}(\delta t^4)$. From the numerical point of view, the most expensive part in RK4 is the calculations of the four slopes, which in the case of Lindblad master equations, boil down to the multiplications of the Hamiltonian and Lindblad operators on the density matrix from both sides. In this sense, the RK4 method has the same numerical complexities as the Taylor series method.

\section{Tensor-Network Implementation of the Taylor Series Method}\label{TN}

\begin{figure}
\centering
\begin{minipage}[t]{0.6\hsize}
\resizebox{1.0\hsize}{!}{\includegraphics{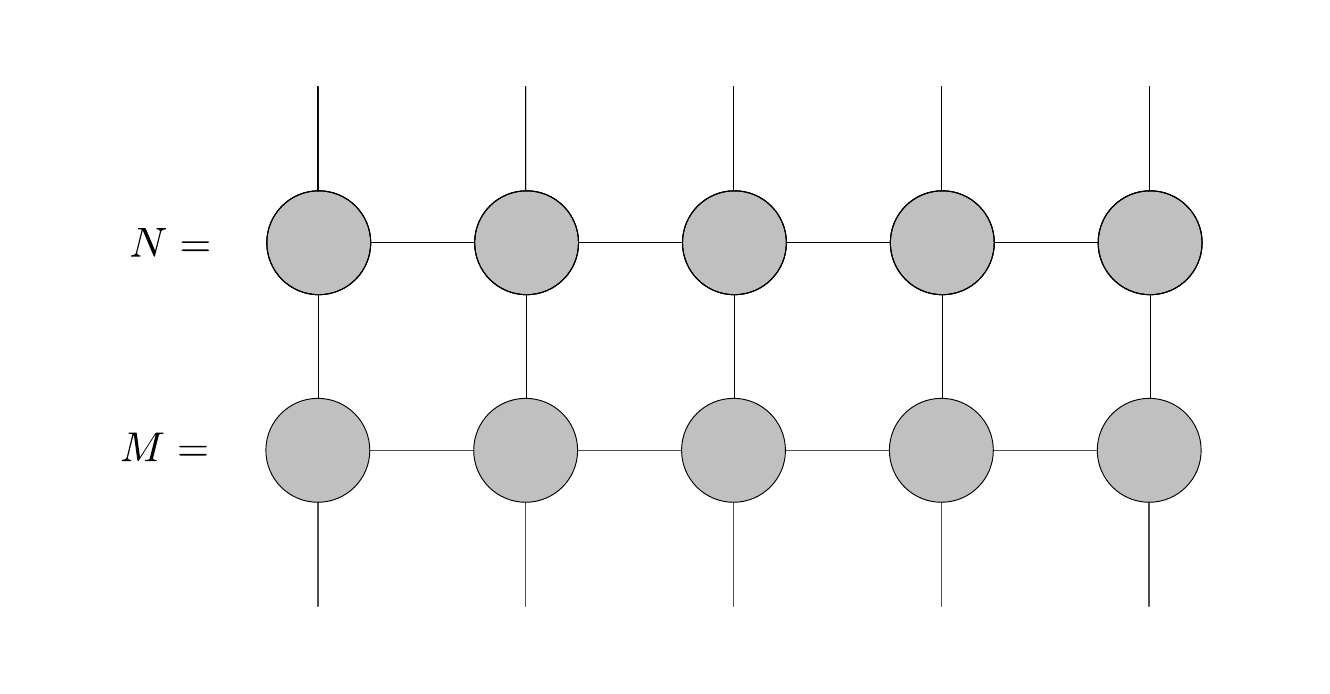}}
\end{minipage}
\caption{Diagrammatic notation of one MPO, $M$, multiplied to another MPO, $N$. The result is a new MPO, $MN$, after contracting the physical indices between them.}
\label{fig_mpompo}
\end{figure}

The procedure in Algorithm~\ref{algorithm} applies in the same way to the tensor-network implementation of the Taylor series method. In this case, the Hamiltonian, Lindblad operators, and the density matrix are all represented as MPOs. The basic operations in Algorithm~\ref{algorithm} are the contraction and addition of MPOs. The former is equivalent to matrix multiplication. Suppose that we have two MPOs with the same physical indices, 
\begin{align}
& M=\sum_{\{i',i\}}{\rm Tr}\left(M_1^{i_1',i_1}M_2^{i_2',i_2}\cdots M_L^{i_L',i_L}\right)\ket{{\bf i}'}\!\bra{{\bf i}} \text{,} \\
& N=\sum_{\{i',i\}}{\rm Tr}\left(N_1^{i_1',i_1}N_2^{i_2',i_2}\cdots N_L^{i_L',i_L}\right)\ket{{\bf i}'}\!\bra{{\bf i}} \text{,}
\end{align}
where ${\bf i}$ and ${\bf i}'$ are the shorthand notations for physical indices $\{i_1,i_2\cdots,i_L\}$ and $\{i'_1,i'_2\cdots,i'_L\}$ with prime levels $0$ and $1$, respectively. When contracting $M$ with $N$, we first raise the prime level of all physical indices of $M$ by $1$ to obtain
\begin{align}
M'=\sum_{\{i'',i'\}}{\rm Tr}\left(M_1^{i_1'',i_1'}M_2^{i_2'',i_2'}\cdots M_L^{i_L'',i_L'}\right)\ket{{\bf i}''}\!\bra{{\bf i}'} \text{.}
\end{align}
In this way, the bra indices of $M'$ and the ket indices of $N$ have the same prime level. Then, the tensor-network contractions are performed per site between the same indices with the same prime level. After that, the indices with prime level $2$ are lowered back to $1$, so that the resulting MPO has pairs of physical indices with prime levels of $0$ and $1$. As a consequence, the dimensions of bond indices of the resulting MPO are increased, becoming the products of the corresponding dimensions of the bond indices from $M$ and $N$. Figure~\ref{fig_mpompo} shows the graphical illustration of the contraction of two MPOs. The addition of $M$ and $N$ is performed as follows. For the $k$-th site, combine the tensors $M_k^{i_k,i'_k}$ and $N_k^{i_k,i'_k}$ into a single tensor $S_k^{i_k,i'_k}$. This is achieved by block diagonal concatenation, where bond dimensions are increased to accommodate both $M$ and $N$,
\begin{align}
S_k^{i_k,i'_k}=\begin{pmatrix}
M_k^{i_k,i'_k} & 0 \\
0 & N_k^{i_k,i'_k}
\end{pmatrix} \text{.}
\end{align}
The inflation of the dimensions of bond indices of the resulting MPO from both contraction and addition can be well controlled by the singular value decompositions (SVDs). With the \textsc{julia} library ITensor~\cite{Fishman_SciPostPhysCodeb_2022}, the contraction of two MPOs can be conveniently done by a single code line \texttt{MN=ITensors.apply(M, N)} without manually handling the prime levels of indices. The addition of two MPOs can be intuitively performed with the plus sign, \texttt{S=M+N}.

\section{Minimally Entangled Typical Thermal States}\label{METTS}

\begin{figure}
\centering
\begin{minipage}[t]{0.5\hsize}
\resizebox{1.0\hsize}{!}{\includegraphics{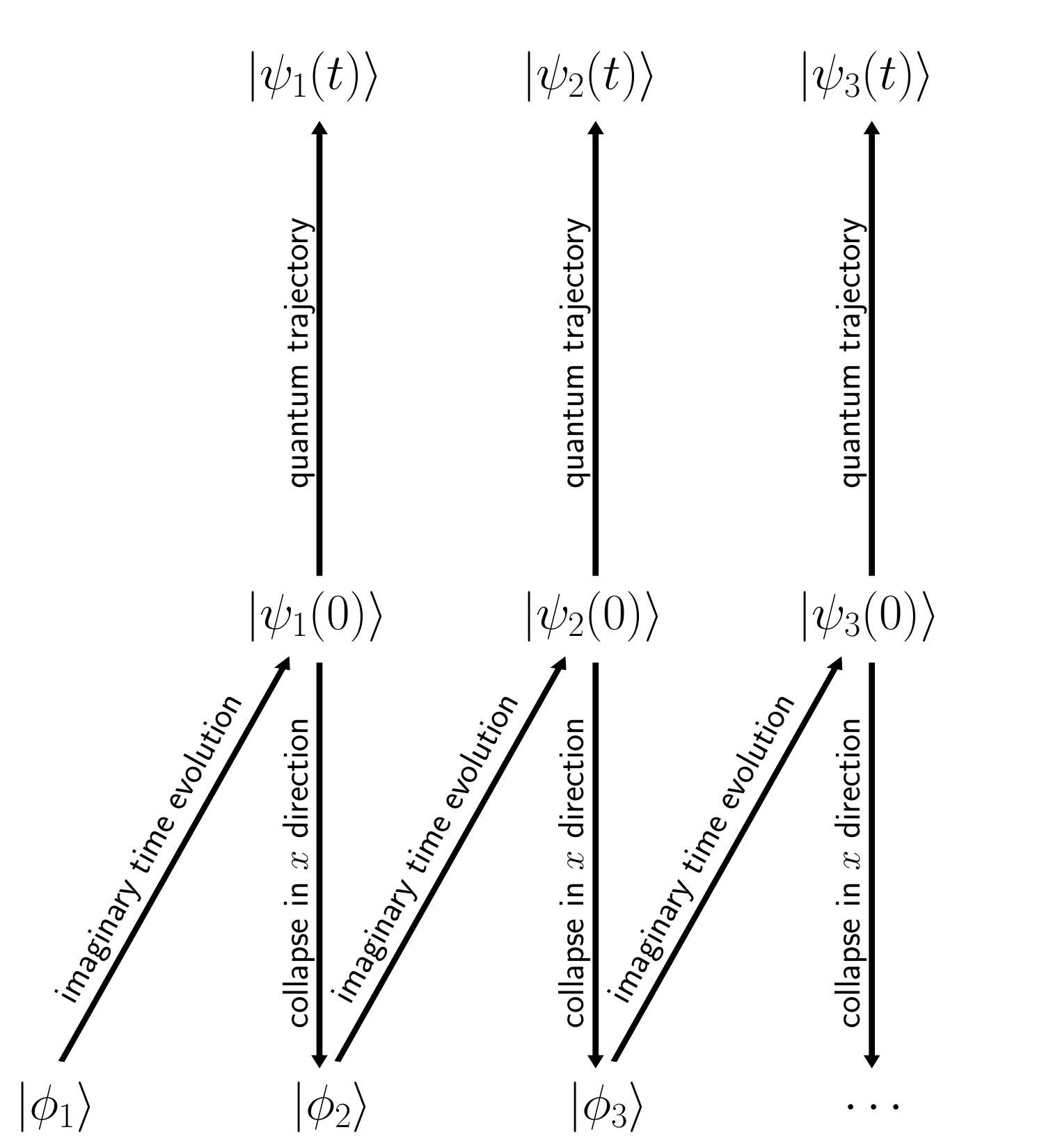}}
\end{minipage}
\caption{Schematic illustration of the METTS algorithm for generating a set of typical states $\{\ket{\psi_i(0)}\}_{i=1}^N$ that represent the initial thermal equilibrium state for the quantum trajectories of the Heisenberg chain. $\{\ket{\phi_i}\}_{i=1}^N$ are the collapsed product states in the $x$ direction. The initial thermal equilibrium state is approximately given by $(1/N)\sum_{i=1}^N\ket{\psi_i(0)}\!\bra{\psi_i(0)}$, where $N$ is the number of typical states (METTS).}
\label{fig_metts}
\end{figure}

The stochastic trajectories for the Heisenberg chain are simulated starting from a thermal equilibrium state. For this purpose, we need to unravel this initial thermal state into a set of pure states. This can be achieved with an algorithm called minimally entangled typical thermal states (METTS)~\cite{White_PhysRevLett_2009, Stoudenmire_NewJPhys_2010}. This is a finite-temperature algorithm for generating a set of typical states representing the Gibbs canonical ensemble. For the Heisenberg chain, we can generate a typical state called metts $\ket{\psi_i(0)}$ from a product state $\ket{\phi_i}$ by
\begin{align}
\ket{\psi_i(0)}=\frac{1}{\sqrt{{\cal P}(\phi_i)}}{\rm e}^{-\beta H/2}\ket{\phi_i} \text{,}
\end{align}
where ${\cal P}(\phi_i)=\braket{\phi_i|{\rm e}^{-\beta H}|\phi_i}$. Here, the $0$ in $\ket{\psi_i(0)}$ means the starting time $t=0$ for quantum trajectories. The imaginary time evolution can be realized with TEBD or TDVP. The thermal equilibrium state can then be represented as a number of metts
\begin{align}
\rho_{\rm eq}(0)=\frac{{\rm e}^{-\beta H}}{Z}=\sum_{\{\phi_i\}}\frac{{\cal P}(\phi_i)}{Z}\ket{\psi_i(0)}\!\bra{\psi_i(0)} \text{,}
\end{align}
where $Z$ denotes the partition function, and ${\cal P}(\phi_i)/Z$ is therefore the weight of the corresponding metts $\ket{\psi_i(0)}$. A Markov chain of the product states $\ket{\phi_i}$ can be constructed so that the corresponding metts $\ket{\psi_i(0)}$ distributes according to the probability distribution ${\cal P}(\phi_i)/Z$. The METTS algorithm consists of the following steps:
\begin{itemize}
\item obtaining a metts $\ket{\psi_i(0)}$ from a product state $\ket{\phi_i}$;
\item collapsing the metts $\ket{\psi_i(0)}$ into a new product state $\ket{\phi_{i+1}}$ with the probability ${\cal P}(\phi_i\to\phi_{i+1})=|\braket{\phi_{i+1}|\psi_i(0)}|^2$;
\item repeating the above procedure with the new collapsed product state.
\end{itemize}
The schematic illustration of the algorithm is presented in Figure~\ref{fig_metts}. Each new metts $\ket{\psi_i(0)}$ acts as a starting pure state for subsequent stochastic evolution of the quantum trajectories. Using the METTS algorithm, we can successively generate an ensemble of metts $\{\ket{\psi_i(0)}\}_{i=1}^N$. Considering that the occurrence frequency of the metts obtained from the product state $\ket{\phi_i}$ is asymptotically equal to ${\cal P}(\phi_i)/Z$ as $N\to\infty$, the initial thermal equilibrium state can be alternatively expressed as
\begin{align} 
\rho_{\rm eq}(0)\approx\frac{1}{N}\sum_{i=1}^N\ket{\psi_i(0)}\!\bra{\psi_i(0)} \text{,} \label{eq_rho0}
\end{align}
which has the same form of Eq.~(\ref{eq_rhot}). This justifies the algorithm illustrated in Figure~\ref{fig_metts} that the number of metts is compatible with the number of quantum trajectories. In this figure, we indicate that the product states are obtained from the collapse in the $x$ direction. This is the not strictly necessary; it can also be the $y$ or $z$ direction for the Heisenberg chain. The reason for the choice of $x$ direction is that in this way the diagonal elements of the density matrix can be evaluated with much higher accuracy, as the density matrix is formulated in the $z$-spin representation in this work. In other words, the collapse in the $x$ direction results in faster convergence in the evaluation of diagonal elements in the $z$-spin representation.

\printbibliography[title={References}]

\end{document}